\documentclass[twocolumn,amsmath,showpacs,amssymb,aps,nofootinbib,floatfix]{revtex4-1}
\usepackage{epsfig,amsmath,amssymb}
\usepackage{dcolumn,color}
\usepackage{bm}
\usepackage{graphicx}

\newcommand{\beq}{\begin{equation}}
\newcommand{\eeq}{\end{equation}}
\newcommand{\eqcomma}{\phantom{A},\phantom{A}}
\newcommand{\order}[1]{ \mathcal{O} \left( #1 \right) }

\begin{document}

\title{Sound waves and vortices in a polarized relativistic fluid}
\author{David Montenegro$^1$, Leonardo Tinti$^2$, Giorgio Torrieri$^1$}
\affiliation{$^1$ IFGW,Unicamp, Campinas, Brasil}
\affiliation{$^2$ The Ohio State University}
\date{December 8,2007}

\begin{abstract}
We extend the effective theory approach to the ideal fluid limit where the polarization of the fluid is non-zero.  After describing and motivating the equations of motion \cite{longpaper}, we expand them around the hydrostatic limit, obtaining the sound wave and vortex degrees of freedom.  We discuss how the presence of polarization affects the stability and causality of the ideal fluid limit.
\end{abstract}

\maketitle
Relativistic hydrodynamics is an area which currently enjoys a high level of both theoretical and phenomenological 
activity \cite{kodama}.  Heavy ion collision experiments have been shown to be well described by hydrodynamic simulations.  The connection of hydrodynamics to microscopic theory, and the development of hydrodynamics as an effective field theory, are however still not well understood.

The direct observation of polarization of $\Lambda$ particles in heavy ion collisions \cite{upsal} and microscopic strongly coupled systems \cite{spintronics}, as well as phenomena such as chiral transport and anomalies \cite{stephanov1,stephanov2,kharzeev,surowka,igor,shocks} has motivated the phenomenological study of vorticiity in relativistic fluids \cite{uspol,wangpol,csernaipol} and its relationship to microscopic spin polarization.

A "realistic" effective theory for describing the relationship between vorticity and polarization in the ideal fluid limit is however still in development.   Vorticity does not emerge in the transport limit, but rather close to the thermodynamic/hydrodynamic regime.   The former was studied within the usual thermodynamic techniques, updated with the inclusion of angular momentum \cite{becattini1,becattini2,becattini3}, but this is not a realistic setup for a strongly coupled dynamical system, where equilibrium is to a good approximation local rather than global.

Local polarization has also been computed for a vorticose fluid from an extension of the Cooper-Frye formula \cite{csernaipol}, but, if such polarization exists at freeze-out, it must exist throughout the evolution of the fluid and backreact on the vorticity as well.


The problem describing this is a conceptual one, since several features associated with fluid dynamics, such as local isotropy in the comoving frame \cite{landau} and the relativistic Kelvin theorem (also referred to as vorticity conservation) \cite{uspol}, are inherently broken by local polarization \cite{becattini1}.  One therefore needs to agree what ideal hydrodynamics in this context means, before developing a hydrodynamics from these principles.
Usual insights from transport theory can also be misleading since polarization is a breakdown of molecular chaos, since it is a two-particle microscopic phase space correlation \cite{longpaper}
.


An approach to describe this is to use effective field theory \cite{nicolis1,torrieri1,torrieri2,torrieri3,glorioso}, but modify the way the gradient expansion is used: Polarization at global equilibtium is proportional to vorticity, which naively should be suppressed byt the Knudsen number, but in fact it survives at global equilibrium and it is necessary to consider it if one wants to assume local equilibium with non vanishing angular momentum (spinning fluid cell). The other gradients, like the shear stress, being vanishing at equilibrium, can be considered as usual related to dissipation. They can be neglected in the limit of ideal fluids.
In \cite{longpaper} we presented the simplest instance (lowest order in gradients and simplest assumption for the Lagrangian density) of such a theory.  We shall briefly introduce it and examine its behavior close to hydrostatic equilibrium.

In this approach perfect fluid without polarization can be described by three fields, representing the comoving ( ``Lagrangian``) coordinate systems $\phi^I$.
Fluidity can be defined via symmetries, namely the invariance  of the Lagrangian under a volume-preserving
 diffeomorphism invariance  \cite{nicolis1,nicolis2,nicolis3}. In general
\begin{equation}
\label{diffeoinv}
L(\phi_I \rightarrow \xi_I(\phi_J)) \rightarrow L \eqcomma 
 \det\left[ \frac{\partial \xi_I}{\partial \phi_J }\right]=1
\end{equation}
  It directly follows \cite{nicolis2,nicolis3} that the lagrangian is of the form, at the lowest order in gradients
\begin{equation}
L=F(b) \eqcomma b=\sqrt{\det_{IJ} \left[B_{IJ}\right]}\eqcomma B_{IJ}=\partial_\mu \phi_I \partial^\mu \phi_J
\end{equation}
The Lagrangian above straight-forwardly yields the general energy momentum tensor whose conservation yields Euler's equations \cite{nicolis1,nicolis2,nicolis3},
\begin{equation}
\partial_\mu T^{\mu \nu}=0 \eqcomma T^{\mu \nu} = (p + e) u^\mu u^\nu - pg^{\mu \nu}
\end{equation}
if one substitutes, for energy density $e$, pressure $p$ and chemical potential  $\mu$
\begin{equation}
e= \frac{dF(b,\mu)}{d\mu} -F(b,\mu) \eqcomma p= F(b,\mu)-\frac{dF(b,\mu)}{db} b 
\end{equation}
the latter, defined in terms of an internal $U(1)$ conserved current $J^\mu$ and the chemical shift symmetry \cite{nicolis3}
\[\  L(\phi_I,e^{i\alpha}) \rightarrow L(\phi_I,e^{i\alpha+\psi(\phi_I)})  \]
to be
\[\  \mu=b^{-1}J^\mu \partial_\mu \psi \]
The lagrangian in general is related to the free energy by a legendre transformation w.r.t. the chemical potential.

The diffeomorphism symmetries and the chemical shift allow for a uniquely defined flow velocity
\begin{equation}
\label{uphidef}
u_\mu = \frac{1}{6 b} \epsilon_{IJK} \epsilon_{\mu \alpha \beta \gamma} \partial^\alpha \phi^I \partial^\beta \phi^J  \partial^\gamma \phi^K = \frac{J^\mu}{\sqrt{-J^\nu J_\nu}}
\end{equation}
with the comoving projector being
\begin{equation}
\label{projector}
\Delta^{\mu \nu} = B_{IJ}^{-1} \partial^\mu \phi^I \partial^\nu \phi^J
\end{equation}
If the system has intrinsic polarization,the coordinates $\phi^I(x)$ are not enough because they do not contain information about polarization.  Moreover, because of the conceptual difficulties described earlier, we need to define precisely what we mean by the ideal hydrodynamic limit, especially because of the polarizations breaking of local isotropy and vorticity conservation.  We take the ideal hydrodynamic limit to mean that dynamics is dependent only on the flow $u^\mu$ and comoving variables, the comoving variables are defined entirely by local entropy maximization, and that only sound waves and vortices arise as local excitations(see Ref. ~\cite{longpaper} for the details. Note that previous attemps to address this issue did not concentrate on this poit \cite{oldspin1,oldspin2}).


Regarding the polarization effective degrees of freedom $y^{\mu\nu}$, they have been defined in Ref~\cite{longpaper} as the (infinitesimal) volume integral of the (space part of the) total angular momentum of a fluid cell in the local comoving frame, normalized using the volume of the cell itself. This can be interpreted as a way to keep track of the angular momentum structure of the fluid during a coarse graining procedure. Angular momentum conservation has always been considered in relativistic hydrodynamics, therefore no new symmetry has to be added, only the new polarization degrees of freedom: the lack of a spin tensor and the symmetry of the stress-energy tensor imply local angular momentum conservation both for ideal and viscous hydrodynamics.

If the polarization were "passively transported along the fluid", in other worlds, if polarization were conserved, then one wold need an additional symmetry on the effective Lagrangian. This situation is similar to the {\it actual} $U(1)$ symmetry considered in Ref~\cite{nicolis3}, where the phase invariance, represented by an explicit dependence of the Lagrangian on the gradient of the local phase, was not enough to recover ideal (non-polarized) hydrodynamics, because the electric current is not guaranteed to be proportional to the energy flux. To enforce this requirement one can introduce the concept of shift symmetry, which can be interpreted as an invariance of physics under local rescaling of the electric charge units. If one uses the same formalism to describe the flow of the spin current ($y^{\mu \nu}$ in the comoving frame), one would have
  

%
\begin{equation}\label{rev_chem_shift}
 L\left(y^{\mu\nu} (x) \right) = L \left( y^{\mu\nu}(x) + f^{\mu\nu}(\phi^I(x)) \right), 
\end{equation}
%
which would represent, not only the invariance of physics under rescaling of the polarization units, but also invariance under an arbitrary addition of an extra polarization which is constant along the flux lines. 
Following \cite{nicolis3} , the invariant that should enter the effective Lagrangian is
$$
\dot y^{\mu\nu} =  u\cdot\partial \, y^{\mu\nu}.
$$
If we apply the ``revised'' chemical shift~(\ref{rev_chem_shift}), $\dot y^{\mu\nu}$ is invariant since   $\partial_\mu \phi_I$ will enter in the derivative and annihillate with $u^\mu\partial_\mu \phi_I$ is vanishing by construction. 

Conservation of angular momentum, even in the ideal limit, does not imply that polarization should correspond to an even approximatel conserved current.  We therefore treat $y^{\mu \nu}$ as an auxiliary field distinct from $\phi_I$, and impose the interaction between orbital and spin angular momentum via the equation of state. In order for local equilibrium to be well-defined, we chose to write directly the polarization degrees as proportional to the local vorticity, 
\begin{equation}
  \label{prevortspin}
  y_{\mu \nu} =\chi(b,\omega^2) \omega_{\mu \nu}
  \end{equation}
a violation of this condition will inevitably result in Goldstone modes and topological constraints which create long-term correlations which make a hydrodynamic limit impossible \cite{longpaper}. 
Note that in an external field such as (but not only) magnetohydrodynamics \cite{kovtun,saso}, this symmetry would be broken and the Lagrangian would depend directly on polarization.


By counting gradients and enforcing symmetries, the lowest order term which respects the  internal diffeomorphism symmetry is $y_{\mu \nu} y^{\mu \nu}$. For example $det[y]$ is a higher term in gradient ($y$ being proportional to the vorticity, which is a gradient), $\epsilon_{\alpha \beta \gamma\rho} \partial^\mu J^\nu y^{\gamma \rho}$ would violate parity and $\partial_\mu J_\nu y^{\mu \nu}$ be higher order in gradient, again.  Parity violating terms would of course be permitted in the context of anomalous hydrodynamics ).  Considering that polarization introduces a correlation between microstates, the presence of polarization at a given entropy $b$ should change the free energy, to leading order in gradient, as $ b \rightarrow b \left(1-c y_{\mu \nu} y^{\mu \nu}  \right)$
where $c$ is a dimensionful constant representing polarizeability (it can be positive for a ferromagnetic material and negative for an antiferromagnetic one).   For dimensional reasons, $c \sim T_0^{2}$
Given this, a physically reasonable way to introduce polarization that we adopted is
\begin{equation}
  \label{rescaledf}
  F(b,y) \rightarrow F\left( b\times f(y)  \right) \eqcomma f(y) = 1-c y_{\mu \nu} y^{\mu \nu} + \order{y^4}
  \end{equation}
For instance, the ideal gas without polarization with  density $n$ of particles of which a fraction $\alpha$ are polarized, assuming nearest-neighbor interactions $b\sim n(1-\alpha^2)$ and $F(b) \sim b^{4/3}(1-\alpha^2)$, making $c \sim 3/4$.


We can now make the link between the lagrangian formulation and usual thermodynamics using the methods of \cite{nicolis3} but with the angular momentum in lieu of chemical potential (note that the collinearity between angular momentum and polarization is what makes this analogy possible).
While physically, because of lack of isotropy, it is non-trivial to relate the usual derivatives of the free energy to what we know as pressure and energy density, the existence of some free energy $F$ to be minimized will lead to a constraint of the type
\begin{eqnarray}
  \label{gd2}
d F (b,y) &=& \partial_b F \, ds + \partial_{y_{\mu\nu}}F \, d y^{\mu\nu} = \\ \nonumber
&=&  -(1-c y^2) F^\prime ds -2cb F^\prime y^{\mu\nu}dy_{\mu\nu}.
\end{eqnarray}
which allows us to link the constant of proportionality in Eq. (\ref{prevortspin}) to $F$ via Legendre transfroms
\[\
 -2cb F^\prime y^{\mu\nu} = -2cb \chi F^\prime \omega^{\mu\nu} \propto \frac{1}{T}
 \]
 Note that this means that the source term $y_{\mu \nu}$ is not a dynamical degree of freedom, since it is fixed by entropy maximization.
 If we had $F$ in terms of a partition function we could find $\chi$ given a local temperature and vorticity explicitly, via a derivation along the lines of \cite{kharzeev} and its similarity with magnetic susceptibility.
 
The equation of motion, using the standard Euler-Lagrange equations, is 
\begin{equation}
\label{EL}
 \partial_\mu J_I^\mu=0,
\end{equation}
where
\[\
 J^\mu_I = 4\, c \, \partial_\nu \left\{ F^\prime \left[ \chi  \left(   \chi + 2 \,  \partial _{\Omega^2}\chi \right) \omega_{\alpha\beta} \, g^{\alpha\{\mu} P_I^{\nu\}\beta} \right]\right\} -
\]
 \[\ - F^{\prime} \left[ u_\rho P^{\rho\mu}_I \left(   1-c y^2 - 2  c  b  \chi  \omega^2 \, \partial_b \chi \right)\right] - 2 c \left(   \chi + 2 \, \omega^2 \, \partial _{\Omega^2} \chi \right)F^{\prime} \times \] 

\[\  \times \left\{ \left[ \chi \, \omega^2 -\frac{1}{b}y_{\rho\sigma} \left(   u_\alpha \partial^\alpha K^\rho - u_\alpha \nabla^\rho K^\alpha \right) \right] P^{\sigma\mu}_I - \right.   \] 
\[\ \left. - \frac{1}{6 b} y_{\rho\sigma}\varepsilon^{\mu\rho\alpha\beta}\epsilon_{IJK}\nabla^\sigma\partial_\alpha \phi^J \partial_\beta \phi^K \right\}.
\]
with $P_{\mu \nu}^K = \partial K_\mu /\partial^\mu \phi^K$, $\nabla^\alpha = \Delta^{\alpha \beta} \partial_\beta$ and $[...],{...}$ corresponding to, respectively, antisymmetrization and symmetrization of the indices, as done in \cite{baier}.

In addition to generally breaking homogeneity andthe relativistic Kelvin theorem, unlike non-polarized hydrodynamics the higher gradient in velocity dependence these equations will be higher than 2nd order in gradient.  To understand the consequences of this, we linearized the hydrostatic limit with "temperature" $T_0 \sim b_0^{1/3}$,
  \begin{equation}
\phi_I = T_0 \left( X_I+\pi_I \right),
  \end{equation}
Note that this is not the actual temperature of the background, it is proportional to that only in the conformal limit.
  Using the notation in \cite{gripaios}, where $\partial \pi$ is shorthand for $\partial_i \pi_J$ and $[\partial \pi]$ for its trace and time derivatives are denoted by dots, the non-polarized hydrodynamics gives the usual wave equation for
  sound waves, the stationary vortex state  polarization terms which will increase the gradients at each order by one unit.  The free part of the equation will be
 \begin{equation}
 \mathcal{L}=  A \left\{ [\partial\pi]  -\frac{1}{2}[\partial\pi\cdot\partial\pi] -\frac{1}{2} \dot\pi^2 \right\} + \end{equation}
\[\ + B \left\{ \vphantom{\frac{}{}} (\partial_\rho \dot\pi)\cdot(\partial^\rho \dot\pi) +  [\partial\dot \pi \cdot \partial \dot \pi]  \right\}
 +\left(\frac{1}{2}A + C\right)[\partial\pi]^2.
\]
where the constants $A,B,C$ are
\begin{equation}
 A = T_0 F^\prime(b_0), \qquad B = A \, c \, \chi^2(b_0,0), \qquad C = \frac{1}{2}b_0^2 F^{\prime\prime}(b_0),
\end{equation} 
giving an equation of motion
\begin{equation}\label{free}
 4B\left\{ \vphantom{\frac{}{}} \ddddot \pi^I - \sum_j \partial_j^2 \ddot \pi^I+ \partial_I \partial_J\ddot\pi^J \right\} + A \ddot \pi^I - 2C\partial_I \partial_J \pi^J 
\end{equation}
   Note that this will {\em always} make these excitations susceptible to Ostrogradski's instabilities, from lowest order \cite{ostro}.

To go further, we decompose the perturbation into an irrotational and rotational parts 
\begin{equation}
  \vec{\pi}=  \vec{\nabla} \varphi (x^\mu) +  \vec{\nabla} \times \vec{\Omega}(x^\mu) 
\end{equation}
, which, in the linearized limit, have their own Fourier modes
\begin{equation}
  \left(
  \begin{array}{c}
    \varphi  \\
    \vec{\Omega}
  \end{array}
  \right) = \int dw d^3 k  \left(
  \begin{array}{c}
    \varphi _0 \\
    \vec{\Omega}_0
  \end{array}
  \right) \exp \left[ i \left(\vec{k_{L,T}}.\vec{x} - w_{L,T} t \right)  \right]
  \end{equation}
For the Longitudinal ones we have
\begin{equation}
  \label{omegel}
 4B \,  w_L^4 - A w^2_L +2C k^2 = 0,
\end{equation}
We note that this differs markedly from the usual effective theory expansion \cite{nicolis1}, where the dispersion relation is of the form $w= \sum_i A_i k^i$.  This difference, instrumental in our conclusions, follows from the fact that the effect of polarization does not follow from a microscopic gradient, but from a ``source`` of a conserved quantity, relativistic angular momentum \cite{longpaper}.  This is defined in a Lorentz-covariant way, and affect the 0th component on the same footing as the spacelike components.

 In order for Eq. \ref{omegel} to have only real solution
\begin{equation}
 A^2 -32 BC \, k^2 \ge 0.
\end{equation}
It is alwas consistent for small momenta, but, depending on the sign of $BC$, stable excitations may disappear at high momenta. 

\begin{figure}
  \epsfig{width=0.45\textwidth,figure=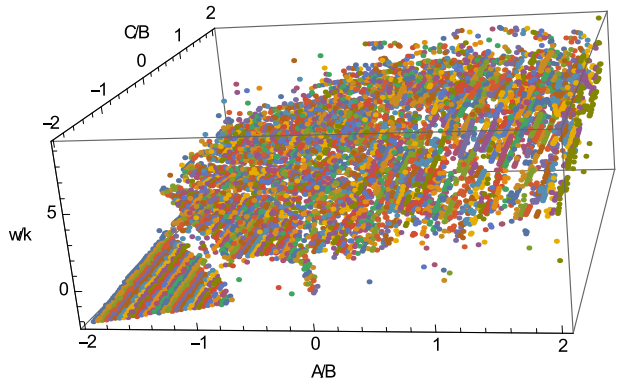}
  \caption{\label{dispersion}The maximum of the soundwave velocity $w/k$ plotted against the parameters $A/B$ and $C/B$. All axes are in logarithm base 10
  Note that we always pick the lowest maximum $w/k$ physical ($\omega>0$ and real) solution of Eq. \ref{omegel} at its maximum.  This generates the noisiness of the plot }
\end{figure}

The two solutions for $w_L^2$ are

\begin{equation}
 w_L^2 = \frac{A}{8B} \pm\sqrt{\left(\frac{A}{8B}\right)^2 - \frac{C k^2}{2 B}}
\end{equation}
Note that, depending on the parameters and on the momentum scale, this can be complex. An imaginary part to the frequency corresponds to damped modes and instabilities.  For the transverse modes we get
\begin{equation}
 4B \, w_T^4 - 4B \, k^2 \, w_T^2 - A \, w_T^2 = 0.
\end{equation}
There is always a couple of non hydrodynamic modes associated withe the vorticous excitations. The solutions for $w_T^2$ are

\begin{equation}
 w_T = 0, \qquad w_T^2 = \frac{A}{4B} + k^2,
\end{equation}
We have calculated the maximum possible wave velocity for $w_L$ numerically, and it is shown as a function of the coefficients in Fig. \ref{dispersion}
On the one hand, as one can see, polarization has important consequences on vortex non-propagation \cite{nicolis1}.  For a general equation of state, a change in temperature due to the propagation of a sound wave will also change the susceptibility.   Because of conservation of angular momentum, this will also change the vorticity.  
This mixing between sound waves and vortices due to the sensitivity of the equation of state to polarization will allow vortices to propagate and create an effective hydrostatic limit even in the hydrodynamic limit in the presence of fluctuations, which could lead to stabilization \cite{longpaper} for small vorticities, the ``fluctuation-driven turbulence'' phase conjectured in \cite{torrieri1,torrieri2}.
On the other hand, the ideal hydrodynamic limit also leads to a non-causal propagation for both sound and vorticity groupvelocity. In Fig. \ref{dispersion}  $A/B$ and $C/B$ are systematically varied, and it is show the highes phase velocity $w/k$ is maximized.  
For all significant values ($\gg 10^{-2}$) of these parameters, the maximun of $w/k$ is very high, and in the same region the group velocity significantly exceed the causal limit, possibly because of a too fast phase velocity of excitations in the region.

This is actually a reasonable result, as it is a direct consequence of the fact that the Free energy $F(b,y)$, and 
hence the local dynamics, is sensitive to an accelleration (similar non-causal dissipative violations of propagation were found at higher gradient order in \cite{denicolvort,baier}).  Ostrogradski's theorem guarantees such Lagrangians are unstable (and one needs
the Lagrangian picture to see this instability), and hence hydrodynamic systems including a response to angular momentum in the EoS will not have a stable ideal hydrodynamic limit.  
It must be noted that removing the ansaztz in~(\ref{rescaledf}), using instead a generic $F(b,y)$ and removing the assumption of $y^{\mu\nu}$ to be space-like, that is using for the right hand side of Eq.~(\ref{prevortspin}) the full antisymmetric part of the gradient of the four velocity instead of the vorticity only, would not change 
  small perturbations over the hydrostatic background, up to the values of the constants $A,B,C$ in Eq.~(\ref{free}). 

One would need then to take into account the polarization as a separate dynamical degree of freedom from the vorticity and include $\dot y^{\mu\nu}$ (and $y^{\mu\nu}$ since polarization is not a conserved quantity). Note that in this way it is possible to obtain an Israel-Stewart type equation
\begin{equation}
\tau_\Omega  u_{\alpha} \partial^\alpha y_{\mu \nu} + y_{\mu \nu} = \chi \omega_{\mu \nu} + \order{\omega^2}
\end{equation}
which could  resolve this issue, with an appropriate relaxation time $\tau_\Omega$, like it did for first order viscous hydrodynamics. The procedure to use is similar to what has already been done in Ref~\cite{torrieri3} to insert a relaxation time equation for the pressure correction, however it is important to remind that the current formalism used is incompatible with dissipative processes (indeed any time irreversible process), and it must be extended as it was done in Ref~\cite{torrieri3} itself. It is mandatory to double the degrees of freedom (or double the time dimension, with the average of the two time coordinates as the physical time) in order to theoretically justify an effective Lagrangian as the leftover of the underlying  (full) Lagrangian, after "integrating out" the microscopic degrees of freedom. In this extended formalism angular momentum (and in fact even four-momentum) conservation has to be enforced on the with additional conditions, fore more informations on the subject see, for instance Ref.~\cite{Galley:2014wla,Galley:2012hx}.

This is very different from ideal spinless hydrodynamics, where Navier-Stokes equations necessitate of relaxation dynamics but Euler equations are well-defined.
In hydrodynamics with polarization, there seems to be a general conflict between causality and the non-dissipative regime, one which might lead to a quantitative lower limit for dissipation (the $
\tau_\Omega$ needed to restore causality) in systems whose fundamental constituents have spin.  This will be quantitatively examined in a subsequent work.

In conclusion, we took the linearized effective theory for ideal hydrodynamics in the presence of polarization and examined the behavior of the lowest-lying modes, sound wave and vortical.
Unlike in normal ideal hydrodynamics, the two mix.   On the one hand, this is likely to stabilize the vortex mode by adding an effective soft energy gap to it.
On the other, causality of sound propagation is no longer guaranteed even in the ideal limit unless polarization effects are soft enough.  Fixing this most likely requires introducing relaxation dynamics already from the ideal fluid limit, and this has implications for the minimum viscosity for fluids whose microscopic degrees of freedom have non-zero spin.

\section*{Acknowledgements}
\textit{Acknowledgements} GT acknowledges support from FAPESP proc. 2014/13120-7 and CNPQ bolsa de 
produtividade 301996/2014-8. LT was 
supported by  the  U.S.  Department  of  Energy,  Office  of  
Science,  Office of Nuclear Physics under Award No. DE-SC0004286   
and  Polish National Science Center Grant DEC-2012/06/A/ST2/00390. DM would like to acknowledge CNPQ graduate fellowship n. 147435/2014-5
Parts of this work were done when LT visited Campinas on FAEPEX fellowship number 2020/16, as well as when GT participated in the INT workshop "Exploring the QCD Phase Diagram through Energy Scans
"  We thank FAEPEX and the INT organizers for the support provided.
This work is a part of the project INCT-FNA Proc. No. 464898/2014-5.\\
We wish to thank Miklos Gyulassy for enlightening discussions which posed the conceptual challenges that eventually led to this work,  and Mike Lisa for showing us experimental literature and useful discussions.


\end{document}